\documentclass[a4paper]{article}
\usepackage[numbers]{natbib}
\usepackage{graphicx}
\usepackage{url}
\usepackage{hyperref}
\usepackage[usenames,dvipsnames]{xcolor}
\usepackage{amsmath,amsthm,amssymb,amsfonts,natbib,
	hyperref, color, graphicx,ulem, float,wrapfig,cancel}
\usepackage{natbib}
\usepackage{bm}
\usepackage{url}
\usepackage{titling}
\usepackage{tikz}
\usetikzlibrary{positioning}
\usepackage{verbatim}
\usetikzlibrary{trees}
\usetikzlibrary{positioning}

\begin{document}
\bibliographystyle{plain}


\newcommand{\dd}[1]{\mathrm{d}#1}

\title{Enhancing path-integral approximation for non-linear diffusion with neural network}

\author{Anna Knezevic \\ 
 }

\date{%
    Curtin School of Elec Eng, Comp and Math Sci (EECMS)\\%
        \today
}

\begin{titlingpage}
    \maketitle
    
\abstract{Enhancing the existing solution for pricing of fixed income instruments within Black-Karasinski model structure, with neural network at various parameterisation points to demonstrate that the method is able to achieve superior outcomes for multiple calibrations across extended projection horizons. 
}

\subsection*{Key words} 
Short rate model, Interest rate model, intractable model, distribution, machine learning, artificial neural network.

\end{titlingpage}

\section{Introduction}
In order to approach the understanding of randomness in context fixed income markets a large number of models have been developed. Among those, short rate models have the advantage of simplicity, and kernel of truth in the approach to uncertainty that exists around the pricing process. Log-normal models create a volatility structure that is tied to the current level of interest rates, which creates a framework for pricing instruments that is more aligned with level-linked volatility.

\section{Background- literature review}
Short rate models have been a popular choice of nominal interest rate or credit spread models\cite{buhler1999empirical}. Expanded version of these models has a number of advantages covered in other texts discussing the necessity of expanding stochastic models beyond short rate factor \cite{two-factor}\cite{turfus2019two}, log-normality of these models permits volatility that is proportional to level of interest rate \cite{lognormprop}, and non-negative interest rate conditions, which were useful until most recent past, when model was adjusted to accommodate for negative interest rates by including displacement\cite{Moodys}. However, the log-normal formulation of short rate models also presents challenges due to absence of analytic formulas for financial instruments (e.g. Zero Coupon Bonds - ZCB) \cite{realdon2016tests}.

This paper considers the seminal Black-Karasinski model\cite{black1991bond} and its dynamics for instruments with higher duration. Due to the complexity of this problem recent years have produced a number of excellent papers on the topic including Turfus \cite{turfus2019exact} focusing on performance in low interest rate environment, and overall review on the current work on the model \citep{turfus2021black}, and finally consideration of the marginal distribution over the incremental time step by Stehlnikova\citep{stehlikova2014effective}, and Capriotti\cite{capriotti2020path}.
Due to unbounded upper limit of Black-Karasinski model (expectations of short rate tending to infinity) precision of analytical solutions starts to drop at longer time horizons. The classical evaluation of path dependent instruments (e.g. ZCB) based on these models is performed by binomial trees\cite{brigo2007interest} (this approach is based on recombining trees and a corridor bounds for the results). Recent work in this space includes Antonov and Spector\citep{antonov2011general} and Turfus\cite{turfus2019closed} restricting the change of the rate to a particular value, whereas the work of Capriotti \cite{capriotti2007closed} and Stehlíkova \cite{stehlikova2014effective} approximates the incremental distribution of each time-step.  These approaches all suffer from losing accuracy when volatility increases, either due to calibration or increase in projection horizon.

\section{Structure of the paper}
The existing approximations provide a robust structure for pricing the instruments with shorter maturity. Testing the approximation across various calibrations and horizon points, key weaknesses are identified. Subsequently, by combining the existing methodology with polynomial expansion of key parameter an improvement to the existing formulation for optimisation is attempted.

\section{Existing GTFK approximation}

In recent work\cite{capriotti2020path} uses the idea of path integrals as drivers of stochastic processes in order to apply GTFK\verb||\footnote{ Giachetti and Tognetti \cite{giachetti1985variational} and Feynman Kleinert\cite{feynman1986effective}} approximation to the existing Black-Karasinski problem. The approach has a strong intuitive underpinnings: the average paths of underlying stochastic short rate process should create a robust approximation of the overall dynamics of the problem.
the dispersion coefficient $\alpha$ and $\omega^2$ are created by finding the roots of the equation \begin{equation}
f( \omega)=\omega^{2}-a^{2}-\lambda \sigma^{2}e^{\alpha/2}e^{\bar{x}}
\end{equation}

and subsequently optimising for joint values of the two variables.  As noted in the original work the dependency of these variables on $\bar{x}$ poses the need for a different structure of approximation at higher maturities. The optimisation itself relies on second derivatives, as per most standard approaches, which is an obstacle to the standard setup of neural network models.

However, after analysis calibration possibilities, the equations accuracy exhibits higher sensitivity to higher speed of mean reversion and higher volatility (i.e. the accuracy is lower when these parameters are higher), resulting in less precise estimation leading to pricing errors. This issue is further exacerbated by the long time horizon error. 

Within the current estimation $\bar{x}$ range remains stable across the horizon of the approximation. Yet, perhaps this is not entirely true for every interaction of mean reversion and volatility (the key parameters that drive the accuracy).

\begin{figure}[htp]
\caption{Average sensitivity of error relative to simulations, across non-training calibrations pre optimisation}
\centering
\includegraphics[width=.3\textwidth]{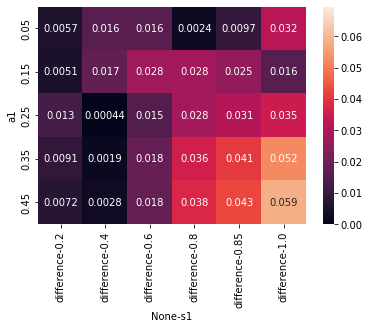}\hfill
\includegraphics[width=.3\textwidth]{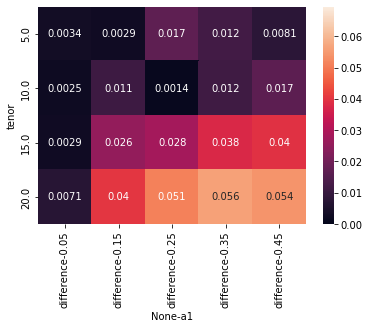}\hfill
\includegraphics[width=.3\textwidth]{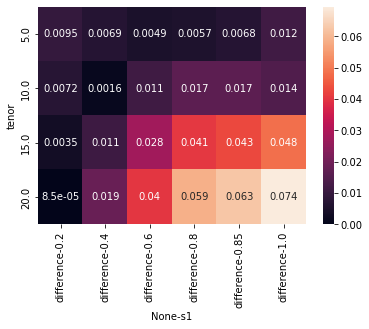}

\label{fig:figure3}

\end{figure}

Both $\alpha$ and $\omega^2$ are equations of exponential function, which is intuitive given the scope of the log-normal models. Acknowledging that some of the errors above can be attributed to simulation noise, the gradation of the error serves as a evidence to the weakness of existing approach.

\subsection{Basic definitions}
 In order to create a neural network capable of replicating the distribution and subsequently the range of ZCB price over the time steps of Black-Karasinski model, a reliable estimator for the distribution of the interest rates is needed. This estimator can then be adjusted in order to find additional terms in key parameters that would help the estimation process, by utilising the Neural Networks in order to approximate the coefficients of the said Taylor series\cite{balduzzi2017neural}.
 
 As the terms of multiplication are constructed based on the constants of calibration  these will be fed as inputs to Neural network, along with the previous value of the process as input. The idea would be to produce coefficients for vector multiplication of Taylor series. 
$$ e^{x}=A_{0}+A_{1}x+\frac{A_{2}x^{2}}{2!}+...+\frac{A_{n}x^{n}}{n!}+...$$

This series was chosen due to the suggestive setup of the problem of integration of exponent of an exponential function.

\section{Model formulation - Neural Network}

Relying on the ability of Neural Networks\cite{wray1995neural} to approximate a large variety of functions, in order to improve the fit of two key variables in the long term structure of interest rates in log-normal model. The neural network needs to be integrated into the model in order to be able to find the new function of $\bar{x}$. The obstacle to this solution is the technical limitations of the neural networks: reliance on derivatives of the functions for optimisation (in reference to the integration over integration along the equivalence class of the path integral), lack of ability to optimise solutions within the optimisation  (in reference to the optimisation to find $\alpha$ and $\omega$).

The hypothesis is that the convergence across $\alpha$ and $\omega$ is missing a more complex structure in equation 1 for $\bar{x}$, that would produce the different roots, that but would be better optimised for both higher and lower volatility. By attempting to examine the influence on those terms further information about potential structure of the equation could be derived.

 \subsection{Neural Network Specification}
Starting from existing formulation of the GTFK Neural Network is created in order to attempt to improve the existing function by creating a polynomial expansion that joins the interdependent functions of $\omega$ and $\alpha$. This result is also fed through to the rest of the computations in order to insure consistency with the original framework.

The input parameters of convergence speed, duration of the ZCB, long-term target interest rate and sigma are used to derive an approximation to ZCB price for a variety of scenario across the input calibration parameters. However, since the error is related to those parameter, the normalisation of the parameters is crucial before attempting to use the input data. The normalisation was performed relative to the entire calibration space.
\tikzset{%
  every neuron/.style={
    circle,
    draw,
    minimum size=1cm
  },
  neuron missing/.style={
    draw=none, 
    scale=4,
    text height=0.333cm,
    execute at begin node=\color{black}$\vdots$
  },
}

\begin{tikzpicture}[x=1.5cm, y=1.5cm, >=stealth]

\foreach \m/\l [count=\y] in {1,2,3,missing,4}
  \node [every neuron/.try, neuron \m/.try] (input-\m) at (0,2.5-\y) {};

\foreach \m [count=\y] in {1,missing,2}
  \node [every neuron/.try, neuron \m/.try ] (hidden-\m) at (2,2-\y*1.25) {};

\foreach \m [count=\y] in {1,missing,2}
  \node [every neuron/.try, neuron \m/.try ] (output-\m) at (4,1.5-\y) {};

\foreach \l [count=\i] in {a,$\theta$,$\sigma$,tenor}
  \draw [<-] (input-\i) -- ++(-1,0)
    node [above, midway] {\l$_{i}$};

\foreach \l [count=\i] in {1,n}
  \node [above] at (hidden-\i.north) {$H_\l$};

\foreach \l [count=\i] in {0,n}
  \draw [->] (output-\i) -- ++(1,0)
    node [above, midway] {$A_{i,\l}$};

\foreach \i in {1,...,4}
  \foreach \j in {1,...,2}
    \draw [->] (input-\i) -- (hidden-\j);

\foreach \i in {1,...,2}
  \foreach \j in {1,...,2}
    \draw [->] (hidden-\i) -- (output-\j);

\foreach \l [count=\x from 0] in {Input, Hidden, Output}
  \node [align=center, above] at (\x*2,2) {\l \\ layer};

\end{tikzpicture}

    Where $n$ is the expansion of $\bar{x}$ to the nth term, and $i$ is the observation of parameterisation and tenor.
 The GTFK approximation due to Capriotti\cite{capriotti2020path} used a joint formulation to find roots of key drivers of Arrow-Debreu densities equation and then relying on assumption drift term being and affine function, derive the equation for zero coupon bond prices and derivatives. Inverting this structure to instead answer question of how much could be accumulated on a single unit over the time horizon, (this can be re-inverted to return to ZCB). In order to correctly reflect this change the drift potential of the Black-Karasinki equation is updated to:
 $$
 V_{BK}(\bar{x})=\frac{a^{2}(b-\bar{x})^{2}}{2\sigma^{2}}-\frac{a}{2}-e^{\bar{x}}
 $$
 This structure allows for a more solvable solutions from perspective of gradient optimisation.

 Additional problem is presented in the integration along the equivalence class of the path integral: within existing Machine Learning setup integrals are not easily reconstructed. Hence a simplification with resampling using the trapezoid rule is applied in its place.

Creating a more complex structure with multiple layers removes the possibility of optimisation via the first derivative.
The normalised coefficients $norm(coeff)$ of each trial calibration (note normalisation values, as well as activation layer weights example is in Appendix 1), are then passed to the dense layer in the following equation:
$$A_{0,...,n}=tanh(H_{1,...,n} \cdot norm(coeff) +bias)$$
$$\dot{\bar{x}}=\bar{x}+ \sum_{i=0}^{n} A_{i}\bar{x}^{i}$$

As mentioned in the Basic Definitions, this can be further improved by overlaying Taylor expansion of exponential with $A_{i}$. The structure of the proposed equation focuses on finding solutions in the nearest region, based on the changes to parameterisation of the model and duration of the instrument simulation dynamics.

The Neural Network has to be designed to take into account that the propagation algorithms rely on approximating the first derivatives of the problem: this means that when dealing with exponential the optimisation has a tendency to go to infinity, which prohibits the algorithm from finding the optimal solution. The usual approach is to restrict the kernel of the activation, clip the change in gradients of the optimiser both on absolute and relative values. However, these measures are ineffectual, since the proposed amendment to the approach focuses on calibration of coefficients within the exponent, increasing non-zero values of $\bar{x}$. The existing solution serves as a basis for finding additional coefficients, to localise the optimisation problem. That implies that the proposed coefficients (that are based on normalised value of the calibration) need to be reduced to a fraction in order not to create exploding gradients problem. 

In order to further improve the optimisation chances the sensitivity of derivatives of subsequent powers of $\bar{x}$ are penalised by applying an additional coefficients as per Taylor expansion (discussed in Basic Definitions). However, use of Taylor expansion is non-essential.

Note, that due to requirement for the input to be normalised in order to be able to be able to converge the neural network and find appropriate additional coefficient for the Taylor expansion polynomial the result of the dense output to the power of exponential is used, since the expected average of the output layer is zero.

Additional measures include careful selection of activation units since usage of sigmoid, linear and the like activation quickly leads to exploding gradients.

Due to exploding gradients and the magnitude of error (1\% relative to large number of inputs) only L1 loss could be used. 

\begin{figure}[htp]
\caption{Average sensitivity of error relative to simulations, across non-training calibrations post optimisation}
\centering
\includegraphics[width=.3\textwidth]{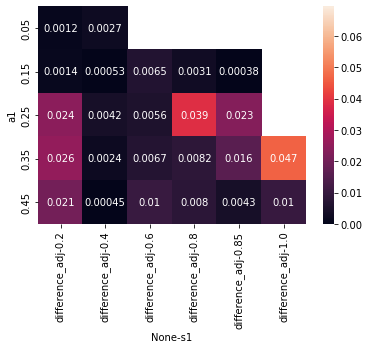}\hfill
\includegraphics[width=.3\textwidth]{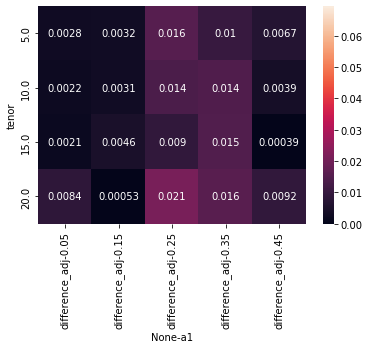}\hfill
\includegraphics[width=.3\textwidth]{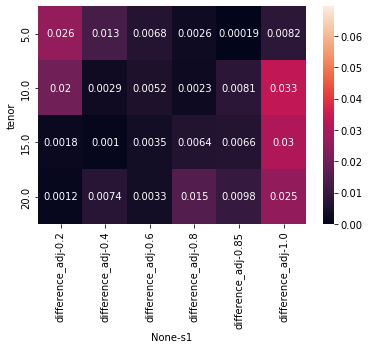}

\label{fig:figure3}

\end{figure}
The density of the predictions are not uniformly distributed with majority of the "exceeding" predictions lying on diagonal of mean reversion speed vs volatility.
\begin{figure}[htp]
\caption{Density of the improved predictions along $\alpha$ (y-axis) and $\sigma$ (x-axis)}
\centering
\includegraphics[width=0.5\textwidth]{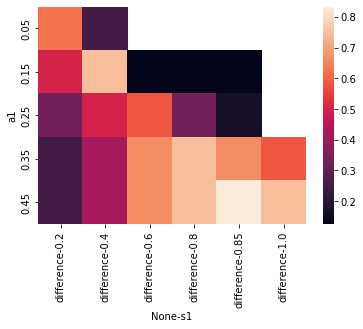}\hfill

\label{fig:figure3}

\end{figure}
The structure of framework for estimation of the expanded polynomial in 
\section{Conclusion}
For a number of subsets, were interaction between $\sigma$ and $a$ are more relaxed, the neural network approach does not produce improvement in results, but with more pronounced interaction between mean reversion parameter and volatility there seems to be additional coefficients in the equation that produce better fit to results for longer horizons.
By calibrating to a large variety of observations and relying on $\alpha$ for sensitivity of the fit to demonstrate that additional coefficients could be applied within the existing structure to create a better fit for higher maturities in calibrations that is further from the original computation space.
Initial ventures into de-coupling of $\omega$ and $\alpha$ parameters did not yield any positive conclusions, restricting the optimisation to only the key sensitivity parameters ($\alpha$, $\omega$, tenor) did not produce additional improvements.
The main improvement came from the use of biases in the optimisation, which suggests that the structure of the existing solution is yet to capture all the effects of the underlying.

\section{Additional findings and further work}
Further analysis on the topic could reveal a sensitivity to extreme levels of theta at higher levels of speed of mean reversion. 
The grid of analysis can be extended further to include higher assumption about long term target of interest rates (range analysed was between 2\% and 6\%), combinations of higher mean reversion and higher sigma combinations (this analysis could be useful for developing economies where the long term projection of short rate would be more accurately reflected with rate in excess of the ones used for this study), and attempting to approximate the unconditional distribution of this process. 
Additional work can be done establishing the times at which the algorithm can be used, vs the original approach.
Additional challenge could be proposed by attempting to extend the existing solution to two factor Black-Karasinski,whereupon the mean reversion target has been promoted to its own stochastic process. 

\clearpage

\section{Appendix 1}

Table of standardisation values
\begin{center}
\begin{tabular}{ c c c}
  & mean & stdev \\ 
 $a$ & 0.2199 & 0.1139 \\ 
  $\sigma$ & 0.6415 & 0.2739  \\ 
 $\theta$ & 0.0469 &  0.0137 \\  
   $r_{0}$ & 0.0401 & 0.0186  \\ 
  Tenor & 5.9707 & 6.6736  

\end{tabular}
\end{center}

Sample activation weights and bias (without Taylor expansion for exponents)
\begin{center}
\begin{tabular}{ c c c c c c c}
  & a & $\sigma$&$\theta$ &  $r_{0}$ & Tenor & bias \\ 
 $A_{0}$ & 2.47E-2 & -1.189E-1& 1.3E-3& 3.99E-2& -9.94E-2&-0.2613\\ 
  $A_{1}$ & -5.528E-1 & 1.771E-1& -3.44E-2& 2.028E-1& -2.822E-1& 0.8864 \\ 
 $A_{2}$ &0 &  -6.9E-3 & 0& 1.81E-2& 5.6E-3&-8.5120\\  
  $A_{3}$& 2.82E-2&-7.620E-2& 0& 1.50E-3& 0&8.6874 \\ 
 $A_{4}$ & 6.719E-1 & -3.589E-1 &4.072E-1& -1.073E-1& -1.532& -0.5019

\end{tabular}
\end{center}

\section{Disclosure of interest} 
There are no competing interests.

\section{Declaration of funding}
No funding was received.

\section{Data availability statement}
Data available on request from the primary author, retention period 7 years from date of publication due to dataset size.
	
\end{document}